\documentclass[useAMS,usenatbib,usegraphicx,times]{mn2e}

\newcommand{\xmm}{{\it{XMM--Newton}}}

\def\gs{\mathrel{\hbox{\rlap{\hbox{\lower4pt\hbox{$\sim$}}}\hbox{$>$}}}}
\def\ls{\mathrel{\hbox{\rlap{\hbox{\lower4pt\hbox{$\sim$}}}\hbox{$<$}}}}

\def\Ms{{\it M}$_\odot$}
\def\kmps{km~s$^{-1}$}
\def\ergps{erg~s$^{-1}$}
\def\kmpspMpc{km~s$^{-1}$~Mpc$^{-1}$}

\def\ergpspsqcm{erg~cm$^{-2}$~s$^{-1}$}

\def\rg{$r_{\rm g}$}
\def\phpspsqcm{ph\thinspace s$^{-1}$\thinspace cm$^{-2}$}

\title[Modulating Fe K emission in NGC3516] 
{Flux and energy modulation of redshifted iron emission in NGC3516: 
  implications for the black hole mass}
\author[Iwasawa, Miniutti \& Fabian]
{K. Iwasawa, G. Miniutti and A.C. Fabian\\
  Institute of Astronomy, Madingley Road, Cambridge CB3 0HA }

\pagerange{\pageref{firstpage}--\pageref{lastpage}}
\pubyear{2004}

\usepackage{times}

\begin{document}

\maketitle

\begin{abstract}
  We report the tentative detection of the modulation of a transient,
  redshifted Fe K$\alpha $ emission feature in the X-ray spectrum of
  the Seyfert galaxy NGC~3516. The detection of the spectral feature
  at 6.1 keV, in addition to a stable 6.4 keV line, has been reported
  previously. We find on re-analysing the XMM-Newton data that the
  feature varies systematically in flux at intervals of 25~ks. The
  peak moves in energy between 5.7 keV and 6.5 keV. The spectral
  evolution of the feature agrees with Fe K emission arising from a
  spot on the accretion disc, illuminated by a corotating flare
  located at a radius of (7--16) \rg, modulated by Doppler and
  gravitational effects as the flare orbits around the black hole.
  Combining the orbital timescale and the location of the orbiting flare,
  the mass of the black hole is estimated to be (1--5)$\times
  10^7$\Ms, which is in good agreement with values obtained from other
  techniques.
\end{abstract}


\section{Introduction}

The Fe K$\alpha$ line is commonly observed in X-ray spectra of Active
Galactic Nuclei (AGN). A narrow line at $6.4$~keV is often seen and
originates most likely from distant material, such as the broad line
regions or the molecular torus. A number of AGN and Galactic Black
Holes Candidates (GBHCs) exhibit broad Fe emission, which is believed
to originate in the innermost part of the accretion flow, where the
line profile is shaped by special and general relativistic effects,
such as Doppler shifts, gravitational redshift and light bending, as a
result of the accreting material moving close to the speed of light
and the large spacetime curvature (e.g., Fabian et al 2000; Reynolds
\& Nowak 2003). While the shape of the relativistic Fe line implies
that the line emitting region is a few \rg ($=GM/c^2$), this does not
constrain the mass of the black hole, which needs a physical unit for
the emitting radius.

In addition to the major line emission around 6.4 keV, transient
emission features at energies lower than 6.4 keV are sometimes
observed in X-ray spectra of AGN. An early example was found in the
ASCA observation of MCG--6-30-15 in 1997, which was interpreted as Fe
K emission induced from a localised flare or from a narrow range of
radii in the inner part of the accretion disc (Iwasawa et al 1999).
More examples followed in recent years with improved sensitivity
provided by XMM-Newton and Chandra X-ray Observatory (Turner et al
2002; Guainazzi et al 2003; Yaqoob et al 2003; Iwasawa et al 2004;
Dov\v{c}iak et al.  2004; Turner et al 2004).  These features can be
attributed to an Fe K$\alpha$ line arising at a particular radius
illuminated by a localised flare. If the illuminated spot is close to
the central black hole, then the line emission is redshifted,
depending on the location of the spot on the disc (Iwasawa et al 1999;
Ruszkowski 2000; Nayakshin \& Kazanas
2001; Dov\v{c}iak et al 2004).

In this context, the flare responsible for the spot illumination is
most likely linked with the accretion disc via magnetic fields and is
entrained with the disc orbital motion. A long-lived flare therefore
orbits the central black hole with a period similar to that of the
accretion disc at the radius where the flare takes place. If the
sensitivity and duration of an X-ray observation are appropriate,
evolution of the emission feature arising from such a spot can be
tracked. When a flare survives more than one orbit, periodic signals
should be observed. Combining this orbital period with the line
emitting radius (in unit of \rg) inferred from the line flux/energy
evolution enables the black hole mass to be derived, as outlined by
Dov\v{c}iak et al (2004).

The Keplerian orbital time around a $10^7$\Ms\ black hole (typical for
Seyfert galaxies) is $\sim 10^4$ s at a radius of 10\rg. Some of the
long XMM-Newton observations last for $\sim 10^5$ s without
interruptions, unlike for low-orbit satellites such as ASCA. Therefore a
few cycles of periodic modulations of line emission induced by an
orbiting spot in a Seyfert galaxy can occur within an XMM-Newton
orbit. However, it is generally assumed that the sensitivity of
currently available X-ray instruments is insufficient for detecting
such orbital motion. Here, we apply an analysis technique, devised to
search for temporal evolution of the Fe K line, to one of the
XMM-Newton datasets in which such a redshifted Fe K emission feature
has been reported.

\section{The XMM-Newton data}

We selected one of the XMM-Newton observations of the bright Seyfert
galaxy NGC3516 (Observation ID: 0107460601), for which Bianchi et al
(2004) reported excess emission at around 6.1 keV in addition to a
stronger 6.4 keV Fe K$\alpha$ line in the time-averaged EPIC spectrum.
The reasons for selecting this dataset as the primary target are as
follows. NGC3516 is relatively bright in the Fe K band and the
detection of the redshifted iron emission feature appears to be more
robust than the other cases (Bianchi et al 2004). There is another
XMM-Newton observation of NGC3516 carried out a few months later, in
which the 6.1 keV feature is not present in the time averaged spectrum
(Turner et al 2002; Miniutti et al 2004), indicating its transient
nature. The black hole mass of a few times $10^7$ \Ms\ has been
estimated from reverberation mapping technique (Ho 1999; Onken et al
2003) for NGC3516. With the black hole mass, a spot in the
relativistic region of the accretion disc can complete a few orbits
within the duration of an XMM-Newton observation (see below) if it
survives, as discussed in the previous section. If the 6.1 keV feature
is produced by such an orbiting spot, associated variability in the
line emission is also observable. Dov\v{c}iak et al (2004) have
already investigated the same dataset by dividing it into three time
intervals of $\sim 27$ ks and found the 'red' feature to be present in
all the three spectra. This implies that the feature lasts at least
for most of the observing duration. If it is variable, as
the spot illumination model would predict, a study at a shorter time
resolution is required. In the other XMM-Newton observation, which is
longer than the one we have selected, although some narrow emission
features have been reported (Turner et al 2002), they are seen only for
a brief period and are fainter than the 6.1 keV feature in the first
observation. We therefore consider the first observation more
promising for a detailed line variability study over the second
observation.

The observation started at 2001 April 10, 11:14 (UT), and the time in
the light curves presented in this paper was measured from this epoch.
We use only EPIC pn data, because of the high sensitivity in the Fe K
band. Single and double events were selected for scientific analysis
presented here. There are time intervals of high background at the
beginning and in the last quarter of the observation. These intervals
were excluded from our analysis, leaving a useful exposure time of 85
ks (of which the live time is 70 per cent due to the Small Window mode
operation of the EPIC pn camera). The level of the background during
the exposure was higher than its typical quiescence and fluctuated by
some degree (see Fig. 1) but has little impact on the results
presented below. The background fraction at the Fe K band is 4.4 per
cent.

The data reduction was carried out using the standard XMM-Newton
analysis package, SAS 6.0, and the FITS manipulation package, FTOOLS
5.3. The energy resolution of the EPIC pn camera at the Fe K band is
$\approx 150 $eV in FWHM.

The time-averaged 2--10 keV observed flux is $2.2\times 10^{-11}$
\ergpspsqcm, and the absorption corrected luminosity is estimated to be
$\sim 0.5\times 10^{43}$\ergps for the source distance of 38 Mpc
($z=0.0088$, $H_0=70$ \kmpspMpc).

\section{Data analysis}


\subsection{Continuum subtraction}

The broad-band X-ray spectrum of NGC3516 is very complex, as a result
of modification by absorption and reflection. Since our interest is on
the behaviour of the relatively narrow feature in the Fe K band, we
designed our analysis method as follows to avoid unnecessary
complication: 1) The energy band is restricted to 5.0--7.1 keV, which
is free from absorption which can affect energies below and above; 2)
the continuum is determined by fitting an absorbed power-law to the
data excluding the line band (6.0--6.6 keV), and is subtracted to
obtain excess emission which is then corrected for the detector
response. When determining the continuum in this way, the spectral
curvature induced by much broad iron line emission (Miniutti, Iwasawa
\& Fabian 2004) is approximated by the effect of absorption and its
contribution subtracted away together with the underlying continuum.


\subsection{The excess emission map on the time--energy plane}

We first investigated the excess emission at resolutions of 5 ks in
time and 100 eV in energy. An image of the excess emission in the
time-energy plane is constructed from individual time-intervals. The
detailed procedure of this method is described in Iwasawa et al
(2004). We have verified that the continuum in each spectrum is
determined reliably and that the line flux measurements are robust
against continuum modelling. The image suggests systematic variations
taking place at the energies of the red feature (5.8--6.2 keV) at
intervals of 25 ks while the 6.4 keV core remains nearly constant,
although this raw digital image is only suggestive due to the low
signal to noise ratio as discussed below.

In general, constructing an image is a useful way to search for events
taking place in two dimensions of interest, e.g., time and energy in
this case, when each event spreads across several pixels.  Conversely,
the events need to be appropriately over-sampled.  In our excess
emission map, the time scale of interest appears to be $\sim 25$ ks,
oversampled by a factor of 5 in time. However, over-sampling can cause
individual pixels to have insufficient statistics. The above
individual 5-ks spectra have $\sim 60$ counts (uncorrected for the
dead time) for the red feature at its peak and $\sim 85$ counts for
the 6.4 keV core. Propagating Poisson error for the continuum
subtraction gives a typical uncertainty for the red feature flux to be
$\sim 40$ per cent. When the individual spectra are investigated
individually, as done by the conventional spectral fitting, those
errors are too large to test for flux variability.

However, as mentioned above, the events of interest take place over
several pixels in the digital image, i.e., the characteristic
variability frequency of the red feature appears to be larger than the
noise frequency, which is equivalent to the time resolution (5 ks).
When this condition is met, low pass filtering is effective in noise
reduction. Therefore we applied weak Gaussian smoothing to the above
image with a circular Gaussian kernel of $\sigma = 0.85$ pixel (10
ks$\times$200 eV in FWHM). This kernel size was chosen so that random
noise between neighbouring pixels is suppressed. This image
filtering brings out the systematic variations of the red feature more
clearly. The light curves of the major line core at 6.4 keV (6.2--6.5
keV) and the red feature (5.8--6.2 keV) are obtained from the filtered
image. Fig.  1 shows the light curves of the excess emission in the
two bands along with the broad-band (0.3--10 keV) light curve. The
error of the line fluxes are estimated from extensive simulations,
which are described in detail in Section 3.4, together with an
assessment of the significance of the line flux variability.


\begin{figure}
\centerline{\includegraphics[width=0.45\textwidth,height=0.4\textwidth,angle=270,keepaspectratio='true']{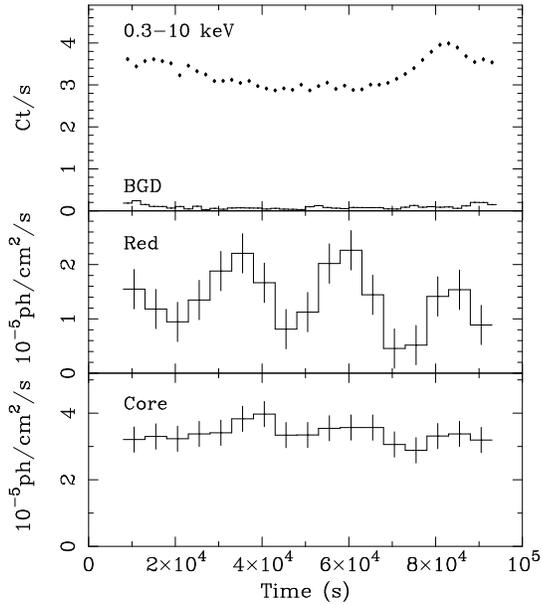}} 
\caption{
  Light curves of the source flux (upper panel), the Fe K red feature
  (middle panel), and the 6.4 keV line core (bottom panel). The source
  flux is measured in count rate in the 0.3--10 keV band with 2-ks
  resolution. The light curve of the background in the same energy
  band, normalised by the extraction area on the detector, is also
  shown in solid histogram. The line fluxes are measured by
  integrating excess emission over 5.8--6.2 keV for the red feature
  and 6.2--6.5 keV for the line core, obtained from the smoothed 5-ks
  resolution image of excess emission in the time-energy plane.}
\end{figure}


\begin{figure}
\centerline{\includegraphics[width=0.24\textwidth,height=0.4\textwidth,angle=270,
    keepaspectratio='true']{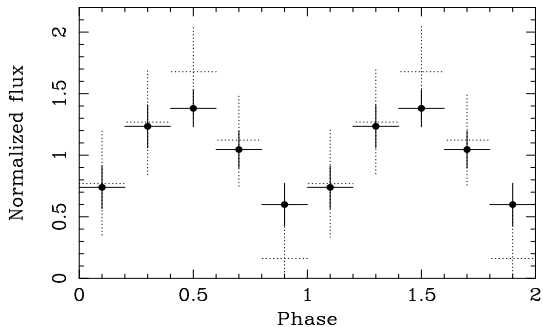}} 
\caption{
  The folded light curve of the red feature. Two cycles are shown for
  clarity. One cycle is 25 ks. The same light curve obtained from
  the unsmoothed data is also shown in dotted line. }
\end{figure}


\subsection{Line flux variability and its characteristic timescale}

The red feature apparently shows a recurrent on-and-off behaviour.
The peaks of its light curve appear at intervals of $\simeq 25$ ks for
nearly four cycles. In contrast, the 6.4 keV line core remains largely
constant, apart from a possible increase delayed from the 'on' phase
in the red feature by a few ks.

Folding the light curve of the red feature strongly suggests a
characteristic interval of 25($\pm 5$) ks. Fig. 2 shows the folded
light curve of the red feature, as well as the one obtained from the
original unsmoothed data of the 5-ks fragments by folding on a 25 ks
interval.  This verifies that the apparent periodic behaviour of the
red feature is not an artifact of the image filtering. However, a mere
four cycles do not secure a periodicity. Whether random noise can
produce spurious periodic variations such as observed at a significant
probability is investigated by simulations below.

\subsection{Error estimate and significance of the red feature variability}


When Gaussian smoothing is applied, the independence of individual
pixels is compromised (not by a great degree in our case of weak
smoothing). This means that simple counting statistics no longer
applies for estimating the error of the flux measurements. We therefore
performed extensive simulations of the line flux measurements, as
detailed below.



In each simulation run, both the 6.4 keV line-core and the red feature
were assumed to stay constant at the flux observed in the time
averaged spectrum throughout seventeen time-intervals. The
normalisation of the power law continuum was set to follow the 0.3--10
keV light curve. From these simulated spectra, an image of the excess
emission in the time--energy plane were obtained and then smoothed,
from which light curves of the two line-bands were extracted, through
exactly the same procedure as applied for the real data. The variance
of each light curve was recorded. This process was repeated 1000
times. 


Since the line flux is kept constant in each simulation, the variance
of the individual light curves represents the measurement uncertainty.
The mean variance of the 1000 simulations was computed and the
standard deviation is adopted as the measurement error. Following this
procedure, the errors of the line fluxes for the red feature and for
the 6.4 keV line core are $0.36\times 10^{-5}$ \phpspsqcm\ and
$0.38\times 10^{-5}$ \phpspsqcm, respectively. With these errors, we
can now assess the significance of the variability in the two
line-bands (Fig. 1).


The $\chi^2$ test shows that a hypothesis of the 6.4 keV core flux
being constant is acceptable. In contrast, the light curve of the red
feature is not consistent with being constant with $\chi^2 = 36.0$ for
16 degrees of freedom. Because of the issue of the pixel independence,
the $\chi^2$ value does not directly translate into a confidence
level. The significance of the red feature variability can instead be
estimated by comparing the $\chi^2$ values against a constant
hypothesis for the real data and the 1000 simulations, or by comparing
the variances directly. We find about 3 per cent of the simulations to
exhibit variability at the same level as the real data, and therefore
conclude that the variability of the red feature is significant at the
97 per cent confidence level.


Assuming now that the red feature is significantly variable, we
examine how likely this is to occur on a 25-ks timescale. Folding the
red feature light curve with the interval of 25 ks (Fig. 2) gives
$\chi^2 = 24.5$ for 4 degrees of freedom. Due to the same limitation
to the statistics mentioned above, we only take the false probability
of 0.1 per cent, which would normally be inferred by the above
$\chi^2$ value, as a target figure for $\chi^2$ tests, and examined
what fraction of the simulated red-feature light curves exhibit
similar level of periodic behaviour for assessing the significance.
The simulated red-feature light curves were folded with six trial
periods between 15 ks and 40 ks at a 5-ks step, and $\chi^2$ values
for the folded light curves were recorded. We find 0.2 per cent show
comparable or larger significance compared to the real data. This is
however limited to the largest trial period of 40 ks. None of the 1000
simulations show comparable periodic signals to the real data for the
trial periods of 35 ks or shorter. The same results were obtained from
direct comparisons of variances of the folded light curves. This is
because the smoothing suppresses the higher frequency (shorter
time-scale) noise and allows the remaining random noise to manifest
itself as spurious periodic variations only on long time scales. The
above test indicates it is unlikely for random noise to produce the
cyclic patterns at the intervals of 25 ks as observed, when our method
is applied as for the real data analysis.


It should however be noted that this test is only against random
independent noise, and does not necessarily prove the 25-ks
periodicity of the line variations. If the line light curve has a
'coloured' (e.g., red) noise spectrum, as AGN continuum emission
usually does, folding tests could give large $\chi^2$ values (e.g.,
Benlloch et al 2001). However, general properties of the temporal
behaviour of iron line emission in AGN are not yet known. Given
the lack of knowledge of the matter, we have to make some assumptions
for the noise spectrum when examining the periodicity. Two limited
cases are presented below.

The power spectrum of the X-ray continuum variability of NGC3516 has
been measured fairly well in a wide frequency range
($10^{-8}$--$10^{-3}$ Hz) and can be approximated by a broken
power-law form (e.g., Markowitz et al 2003). The frequency range of
interest is well above the break frequency at $\approx 2\times
10^{-6}$ Hz and the power spectrum density follows $\propto f^{-2}$,
where $f$ is the Fourier frequency. If the same power spectrum is adopted
for the iron line variability, the r.m.s. variability amplitude
expected on the time scale of 25 ks is 2--3 per cent, much smaller
than the error in line flux measurement. This means that the above
test against random noise is valid in this case.

We further tested a case of the red noise with a large amplitude of
variability using simulated light curves. 1000 light curves were
simulated assuming the same power spectrum slope but a fractional
r.m.s.  amplitude of 35 per cent, similar to that in the red feature
light curve, with the Timmer \& K\"onig (1995) prescription.  Gaussian
noise based on the line flux measurement error (see above) was added
to the light curves. We then folded the simulated light curves with
the six trial periods between 15 ks and 40 ks and compared their
$\chi^2$ values with that for the real data for significance. This
comparison shows less than 2 per cent of the simulated light curves
show stronger periodic signals than the real red feature light curve.
While the above test suggests the probability of the 25 ks periodic
signals being an artifact of red noise is low, given the crude
assumptions, the inferred significance of the periodicity must be
taken as an indication only.  Also, with such a small number of cycles
(and the limited frequency range of the noise spectrum we can
examine), we draw no strong conclusion on the 25 ks periodicity.

\subsection{Line profile variation}

\begin{figure}
\centerline{\includegraphics[width=0.28\textwidth,height=0.41\textwidth,angle=270,
    keepaspectratio='true']{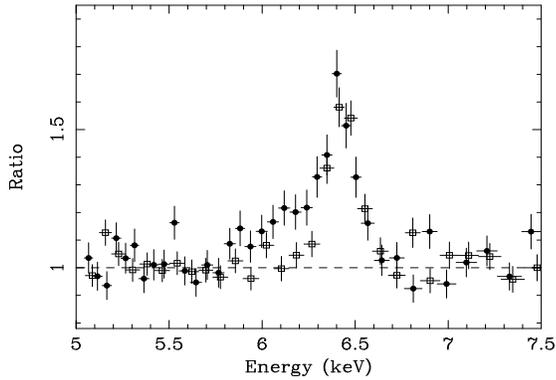}} 
\caption{
The Fe K line profiles during the on (filled circles) and off (open
squares) phases of the red feature. The data are plotted in the form
of a ratio against the best-fitting continuum. The energy scale has been
corrected for the galaxy redshift ($z=0.0088$). }
\end{figure}

Using the above image and light curves as a guide, we constructed two
spectra taken in a periodic manner from the 'on' and 'off' phases to
verify the implied variability in the red feature. The line profiles
obtained from the two spectra are shown in Fig. 3, which can be
modelled with a double Gaussian (Table 1). The 6.4 keV core is resolved
slightly ($\sim 5,000$ \kmps in FWHM) and found in both spectra with
an equivalent width (EW) of 110 eV.  While the 6.4 keV core remains
similar between the two, there is a clear difference in the energy
range of 5.7--6.2 keV due to the presence/absence of the red feature.
In the 'on' phase spectrum, the red feature centred at
$6.13^{+0.10}_{-0.07}$ keV (EW of 65 eV with respect to the continuum
at the centroid energy) is evident. The line flux ratio to the 6.4 keV
line core is $\simeq 0.7$.  It is not detected in the 'off' phase
spectrum. The 90 per cent upper limit of the line flux given in Table
1 is the value when the same line centroid and the width of a Gaussian
as in the 'on' phase spectrum are adopted, and corresponds to $EW<20$
eV. The variability detected between the two spectra is significant at
$4\sigma$. It would be unlikely to detect the variability if time
intervals were chosen arbitrarily, as done by Dov\v{c}iak et al
(2004).

\subsection{Higher time resolution analysis}

Having established the significance of the variability in the red
feature, we now investigate possible time evolution of the feature at
a finer time resolution.  A similar image of excess emission but at a
2 ks time resolution is constructed.  Then elliptical Gaussian
smoothing with dispersion of $1.5\times 1.0$ pixel (7 ks$\times$250 eV
in FWHM) has been applied to obtain the image shown in Fig. 4. The
elliptical Gaussian kernel was chosen so as not to oversmooth the
spectral resolution, which is kept to be 100 eV per pixel in the
original mosaic image. Since the signal-to-noise ratio of the data in
individual time intervals becomes worse, interpretations based on this
image must be treated with cautions. However, further interesting
behaviour of the red feature are noticed (see Fig. 4). The red feature
apparently moves with time during each 'on' phase revealed by the 5-ks
resolution study: the feature emerges at around 5.7 keV, shifts its
peak to higher energies with time, and joins the major line component
at 6.4 keV, where there is marginal evidence for an increase of the
6.4 keV line flux (albeit only suggestive; see also the light curve in
Fig. 1). This evolution appears to be repeated for the on-phases.
\begin{table}
\begin{center}
\caption{
  The 6.4 keV core and the red feature of the Fe K$\alpha $ emission
  in on and off phases. The line energies are measured in the galaxy
  rest frame. The line fluxes are as observed and uncorrected for
  absorption. Errors are of the 90 per cent confidence range for one
  parameter of interest.}
\begin{tabular}{lccll}
& $E$ & {$\sigma$} & \multicolumn{2}{c}{Line flux}\\
& keV & keV & \multicolumn{2}{c}{$10^{-5}$\phpspsqcm}\\[5pt]
Core & $6.42^{+0.02}_{-0.02}$ & $0.045^{+0.020}_{-0.020}$ & 
On: $2.9^{+0.6}_{-0.4}$ & Off: $3.2^{+0.5}_{-0.5}$ \\
Red & $6.13^{+0.10}_{-0.07}$ & $0.15^{+0.25}_{-0.07}$ &
On: $2.1^{+1.3}_{-0.8}$ & Off: $< 0.7$ \\
\end{tabular}
\end{center}
\end{table}

\section{Interpretation and modelling}

The detection of only four cycles is not sufficient to establish any
periodicity at high significance. The 25 ks is however a
natural timescale of a black hole system with a black hole mass of a
few times $10^7$ \Ms, as measured for NGC3516. The finding could
potentially be important and, especially the evolution of the line
emission, warrant a theoretical study. Here we shall assume that the
25 ks timescale is associated with the orbital motion of a corotating
flare which illuminates a spot on the accretion disc, and present
theoretical modelling which reproduces the saw-tooth features seen in
Fig. 4. It also allows us to combine timing and spectral information
for estimating the black hole mass.

Note that since a flare is not expected to last much longer than a few
dynamical timescales nor be strictly periodic due to its orbital drift
as the disc material spirals in (e.g. Karas, Martocchia \& Subr 2001),
the number of detectable cycles will be limited even in much longer
observations. Flares occurring at small radii and/or black holes having
small mass could provide shorter characteristic timescales. However,
detectability is determined by a competition between line intensity
and time resolution at a given throughput of the X-ray telescope. Line
emission from smaller radii is more broadened, which would make it
difficult to distinguish the line emission from the continuum.

\subsection{The corotating flare model}

We adopt a simple model in which a flare is located above an
accretion disc, corotating with it at a fixed radius.  The flare
illuminates an underlying region on the disc (or spot) which produces
a reflection spectrum, including an Fe K$\alpha$ line. The observed
line flux and energy are both phase-dependent quantities and
therefore, if the flare lasts for more than one orbital period, they
will modulate periodically. Details on the expected behaviour can be
found e.g. in the work of Ruszkowski (2000) and Dov\v{c}iak et al
(2004).

The flare orbital period $T$ (equal to the spot period since
corotation is assumed) is related
to the black hole mass ($M_{\rm BH}$) and dimensionless spin ($a$) and to the
radial location $r$ of the flare   
by the following relation (see e.g. Bardeen, Press \& Teukolsky 1972)
\begin{equation}
T = 310~[~a + (~r/r_g~)^{3/2}~]~M_7 
\quad {\mathrm{[\,seconds\,]}}\ ,
\label{eqT}
\end{equation} 
where $M_7$ is the black hole mass in units of $10^7\,M_\odot$.
Hereafter, by radial location $r$ we mean the flare distance from the
black hole axis as measured in the equatorial plane.  It is clear that
if $T$ and $r$ can be inferred from observations, Eq.~(1) provides an
estimate of the black hole mass. Assuming that the observed 25 ks
timescale corresponds to the orbital period, we now model the data and
derive an estimate of $r$ that will constrain $M_{\rm{BH}}$ in
NGC~3516.

\subsection{Constraints on emission radius and black hole mass}

We consider a corotating flare which has a power-law spectrum with
photon index of $\Gamma=1.8$ (typical for Seyfert 1 galaxies) and
emits isotropically at a constant flux in its proper frame. The disc
illumination is computed by integrating the photon geodesics in a Kerr
spacetime from the flare to the disc, and is converted to local
emissivity (see e.g. George \& Fabian 1991). Then, the observed
emission line profile is computed through the widely used ray--tracing
technique (see Miniutti et al 2003 and Miniutti \& Fabian 2004 for
more details).
\begin{figure}
\centerline{\includegraphics[width=0.45\textwidth,angle=0,
    keepaspectratio='true']{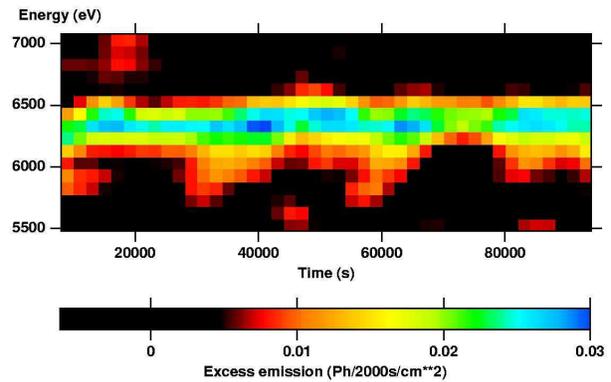}} 
\caption{
The smoothed excess emission map on the time-energy plane. The pixel
size is 2 ks in time and 100 eV in energy. The energy scale is as
observed. The flux of the excess emission is indicated by the false
colour, in the unit of photons/2000-s/cm$^2$. The value of 0.02
corresponds roughly to 10 counts (uncorrected for the dead time due to
the EPIC operation in Small-Window mode). }
\end{figure}
\begin{figure}
\centerline{\includegraphics[width=0.45\textwidth,angle=0,
    keepaspectratio='true']{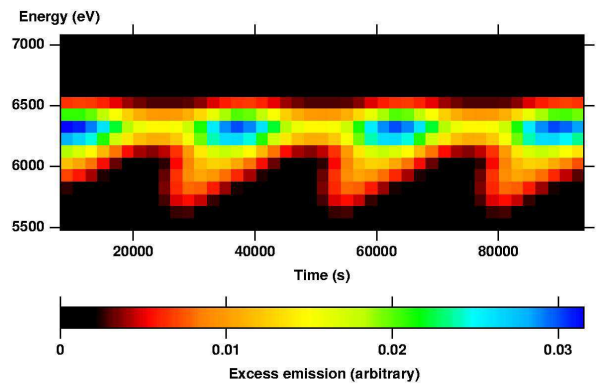}} 
\caption{ The smoothed theoretical time--energy map of the emission
  feature. The time resolution is enforced by dividing the flare orbit
  into $13$ intervals. By assuming an orbital period of $26$~ks
  (consistent with the measured period of $25\pm 5$~ks), each interval
  corresponds to $2$~ks. See
  text for details on the model parameters. No noise has been added to
  the theoretical results.}
\end{figure}

The free parameters of our model are the flare location, specified
by $r$ and the height above the accretion disc ($h$), the
accretion disc inclination $i$, the emission line rest--frame
energy, and the inner disc radius. We assume a maximally rotating
Kerr black hole with spin parameter, $a=0.998$. Note that the
difference in the orbital period between a maximally rotating and a
non-rotating black hole is smaller than the uncertainty of the period
(20 per cent) when $r>3$ \rg, for which our results would be
insensitive to the spin of the black hole. Our assumption on the spin
parameter needs to be verified once a lower limit on the flare radial
position has been obtained.
\begin{figure}
  \centerline{\includegraphics[width=0.32\textwidth,
              height=0.4\textwidth,angle=270,
    keepaspectratio='true']{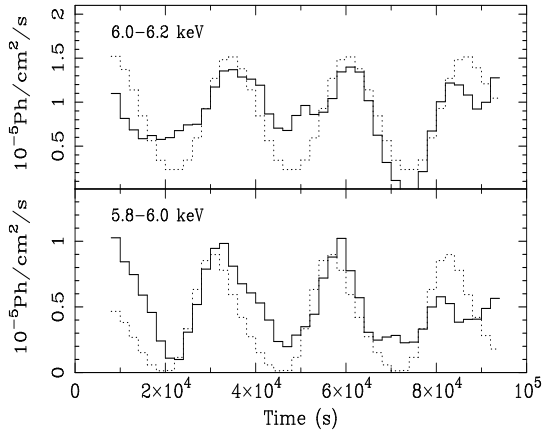}}
\caption{
  A comparison between the observed (solid line) and predicted (dotted
  line) light curves in the 6.0--6.2 keV (upper panel) and 5.8--6.0
  keV (lower panel) bands, obtained from the excess maps in Fig. 4 and
  Fig. 5. This demonstrates the level of match between the two. The
  model describes the amplitudes and delays of the emission feature as
  a function of time reasonably well, given the quality of the data. }
\end{figure}

The variation of energy and flux of the emission feature also depends
on the assumed disc inclination. We have analysed all the BeppoSAX and
XMM-Newton datasets available and found the Fe K line profiles are all
consistent with $i=30^{\circ}\pm 10^{\circ}$. The range of
inclination is in good agreement with previous results based on
the ASCA data (Nandra et al 1997; Nandra et al 1999; \v{C}ade\v{z} et
al 2000).

We have computed the evolution of Fe K emission induced by an
orbiting flare and simulated time-energy maps with the same resolution and
smoothing as in the excess emission map of Fig. 4. A constant, narrow
$6.4$~keV core representing the Fe K$\alpha$ line from distant
material is also added. As pointed out by Dov\v{c}iak et al (2004),
the faint, broad part of the line profile has little contrast against
the continuum and is likely missed in the observed excess map. To
simulate this effect, we set a threshold below which line emission is
assumed to have null intensity. The threshold is an additional
parameter of the model and its adopted range is 30--40 per cent of the
maximum amplitude of the emission feature during the orbital period.

Given the quality of the observed map in Fig. 4 and the simplified
situation assumed in our model, an exact match of the simulated map
with the observation is not to be expected (nor to be looked for).
However, we have explored the parameter space, aiming to reproduce the
key features as outlined below and constrain the maximum range of
$r$ allowed by the data.
The key features we identified are the followings: a) the emission has
a skewed form in the time--energy map (no significant emission is seen
below 6.1 keV during the `off' phase); b) during the `on'
phase, the emission appears at $\sim 5.7$~keV, moves with time to
join the narrow 6.4 keV emission and reaches the maximum energy of
about $6.5$~keV; c) the maximum-to-minimum flux ratio of the feature
in this phase is about 2; and d) the `on' and `off' phases
have approximately equal durations. 

The lack of emission below 6.1 keV during the `off' phase, which makes
the saw-tooth pattern of the excess map (Fig. 4), is the most notable
difference from the predictions in the previous work (Nayakshin \&
Kazanas 2001; Dov\v{c}iak et al 2004). If the spot size is small, the
emission line is always narrow during the orbital period, and the
`off' phase shows a decay both in energy and flux, resulting in an
almost symmetric pattern to the rising `on' phase, especially when the
inclination is $\sim 30^\circ$. In contrast, in our model, the disc
illumination is computed self-consistently and the spot size is
related to the height of the illuminating source (i.e., flare): the
higher the flare, the larger the spot (see also Ruszkowski 2000 for
other relativistic effects mainly due to light bending). This
causes the observed line profile to be broadened.  Therefore when a
flare is located higher than a certain height (typically $h \gs
3~r_g$, depending on $r$), the broadened line emission becomes
undetectable because of the lack of the contrast to the underlying
continuum. This occurs at a phase when the flare is on the near
receding side of the disc, which corresponds to the 'off' phase, and
explains the skewed form of the excess emission in Fig.~4.

Fig. 5 shows one of the theoretical time-energy maps that we consider
to be in reasonable agreement with the observed map because they
reproduce the key features mentioned above. This particular example
assumes a flare with $(\,r\,,\,h\,) = (\,9\,,\,6\,)~r_g$, the
rest-frame line energy of 6.4 keV, a disc inclination of $20^\circ$,
an inner disc radius of 1.24 \rg, and a detection threshold of 37 per
cent.  The normalisation of the excess emission in the theoretical map
is adjusted so that the Fig. 5 can be compared with Fig.  4 directly.
In Fig.~6, light curves in the two selected bands, obtained from the
smoothed observed (Fig. 4) and simulated (Fig. 5) maps are shown to
illustrate the degree of agreement between the model and the observation.

By applying the above procedure we find for the flare radius
\begin{equation}
\label{eqr}
7~r_g \ls r \ls 16~r_g  \ .
\end{equation}
Radii larger than $16~r_g$ are ruled out because 1) the amplitude of
flux variation during the `on' phase becomes too large; 2) the `on'
phase lasts longer than the `off' phase; and 3) substantial emission
below 6.1~keV appears during the `off' phase above the detection
threshold. If the orbital radius is smaller than $7~r_g$, the opposite
trend is seen. Some uncertainty comes from the presence of the narrow
6.4 keV line which blends with the moving feature at energies above
6.2 keV. Note that, since the minimum radius we find is larger than
$3~r_g$, the orbital time is insensitive to our assumption of the
maximally-rotating Kerr spacetime.

Together with $T =(~25\pm 5~)$~ks, and by using Eq. (1),
the flare radius translates to a black hole mass in
NGC~3516:
\begin{equation}
\label{eqM}
1.0 \times 10^7~M_\odot \ls M_{\rm{BH}} 
\ls 5.0 \times 10^7~M_\odot  \ .
\end{equation}

\section{Discussion and Conclusions}

We study the variability of a transient emission feature around 6 keV,
of which detection has been reported previously in the time averaged
X-ray spectrum of NGC3516. The feature appears to vary systematically
both in flux and energy on a characteristic timescale of 25 ks. On
comparing with extensive simulations, the variability is found to be
significant at the 97 per cent confidence level, and the probability
for the variations to occur cyclically as observed purely due to
random noise is very low (see Section 3.4).

The flux and energy evolution of the red feature is consistent with
being Fe K$\alpha$ emission produced by an illuminated spot on the
accretion disc and modulated by Doppler and gravitational effects.
Modelling the observed X-ray data with a relativistic disc illuminated
by a corotating flare above it constrains the radial location of the
flare to be $r = (7-16)~r_g$. This is combined with the
orbital period to provide an estimate of the black hole mass in
NGC~3516 which is $M_{\rm{BH}} = (1-5)\times 10^7~M_\odot$.

One caveat is any weak continuum variation correlated with the line
variation. Since the flare is assumed to corotate with the disc, its
direct emission should also produce modulation in the continuum in
roughly the same way as the red feature. Ideally, this constraint
should be obtained from the ionizing flux of the Fe K line, i.e., the
continuum flux above 7.1 keV, but because of the poor signal to noise
ratio of the light curve in those energies, the 0.3--10 keV band light
curve (Fig. 1) is used instead. If the spectrum of the flare is harder
than the total continuum emission, the limit given below would
increase. The best constraint is obtained from the second and third
'on' phases where the continuum light curve is relatively flat.
Excess flux increases during those periods are of the order of 5 per
cent, which represents an upper limit on the continuum modulation.
Since the line flux of the red feature is a small fraction ($8\pm 5$
per cent on average) of that of the total broad Fe K
emission\footnote{Note that the broad red wing of this line emission
  has been subtracted away in the excess map (see Section 3.1)}, the
relative contribution of the flare, which produces the red feature, to
the total continuum flux is accordingly small, and the line flux ratio
gives an estimate of that.  Based on the estimate, the expected
variability amplitude in the {\it total} light curve due to the
modulation of the orbiting flare is computed for the range of flare
radius and inclination of the disc derived above. It ranges over 3--19 per
cent, which contains the observed limit towards the lower bound. The
amplitude is smaller when the flare radius is larger and the
inclination is smaller. We note the above estimate assumes other X-ray
sources above the disc are entirely static. Other short-lived flares
occuring at different radii may mask the continuum modulation.

The corotaing flare model is perhaps the simplest explanation for the
evolution of the red feature, but there may be other models which do
not invoke accompanied continuum modulation. If the red feature is
related to a non-axisynmetric (e.g., spiral) structure on the disc with
high emissivity, such as that resulted from density and ionzation
perturbations or magnetohydrodynamic turbulance considered by Karas et
al (2001) and Armitage \& Reynolds (2003), respectively, then the line
flux modulates as the structure orbits around but can be independent
from the illuminating source.

Our result on $M_{\rm BH}$ is complementary to those from the
reverberation mapping technique which are based on emission from
clouds much far away from a central hole and unlikely to see those
clouds to complete a whole orbit in a reasonable time. The estimates
of $M_{\rm{BH}}$ in NGC~3516 from reverberation mapping lie in the
range between $1 \times 10^7~M_\odot$ and $4 \times 10^7~M_\odot$. The
most recent result obtained by combining the H$\alpha$ and H$\beta$
emission lines is $(1.68 \pm 0.33) \times 10^7~M_\odot$ (Onken et al.
2003), while the previous analysis based on H$\beta$ alone gave
$2.3\times 10^7~M_\odot$ (Ho 1999; no uncertainty is given). It is
remarkable that our estimate of the black hole mass in NGC~3516 is in
excellent agreement with the above results. Although the systematic
flux and energy variability we report here is only tentative, the
above agreement supports our interpretation. 

Our results indicate that present X-ray missions such as \xmm\ are
close to probing the spacetime geometry in the vicinity of
supermassive black holes if their observational capabilities
are pushed to the limit. Future 
observatories such as {\it{XEUS}} and
{\it{Constellation-X}}, which are planned to have much larger collecting
area at $6$~keV, will be able to exploit this potential and
map the strong field regime of general relativity with great accuracy.

\section*{Acknowledgements}

This research uses the data taken from the XMM-Newton Science Archive
(XSA). We thank Julien Malzac for help with simulating the light
curves used for the siginifcance test. ACF thanks the Royal Society
for support. GM and KI thank PPARC for support. KI also expresses great
 thanks to late his father, who ceased only a few days after the acceptance of this paper, for encouragements.

\end{document}